\documentstyle[prb,aps,floats,epsfig,eqsecnum]{revtex}
\begin{document}
\title{Metamagnetism in  the XXZ-model with next to nearest neighbour coupling}
\author{C. Gerhardt and
K.-H. M\"utter\footnote{e-mail:muetter@wpts0.physik.uni-wuppertal.de}}
\address{Physics Department, University of Wuppertal, 42097 Wuppertal, Germany}
\author{H. Kr\"oger}
\address{Department of Physics, Universit\'e Laval,
Qu\'ebec, Qu\'ebec G1K 7P4, Canada}

\date{\today}
\maketitle
%
%
\begin{abstract}
We investigate groundstate energies and magnetization curves in the
one dimensional  XXZ-model with next to nearest neighbour coupling
$\alpha>0$ and anisotropy $\Delta$ ($-1 \le \Delta \le 1$) at $T=0$. In between
the familiar ferro- and antiferromagnetic phase we find a transition
region -- called metamagnetic phase -- where the magnetization curve
is discontinuous at a critical field $B_c(\alpha,\Delta)$.   
\end{abstract}

\section{Introduction}
Experimental results for the magnetization curves of $Fe_xMn_{1-x}TiO_3$, 
$GdNi_2Sb_2$, $GdCu_2Sb_2$ or $Tb_{1-x}Sc_xMn_2$ show a rapid increase
(or discontinuity) if the applied $B$-field exceeds a critical value
$B_c$. For $B > B_c$ the substance is almost fully magnetized. This
phenomenon is called 'spin-flip' or 'metamagnetic' transition
\cite{ito,berna,kaczmarska}. There have been made
various attempts to explain the 'metamagnetic' transition in the
context of Ising-like Hamiltonians. It is the purpose of this paper to
show that discontinuities in the magnetization curve can be seen as
well in the one dimensional spin-$\frac{1}{2}$ 
XXZ-model with next to nearest neighbour (nnn)  coupling 
\begin{eqnarray}
     H(\alpha,\Delta,B) & = &
    J_1 \sum_{i=1}^N S_i^x S_{i+1}^x + S_i^y S_{i+1}^y + \Delta S_i^z S_{i+1}^z
    \nonumber \\
    & & + J_2 \cdot (S_i^x S_{i+2}^x + S_i^y S_{i+2}^y + \Delta S_i^z
    S_{i+2}^z)  + B S_i^z
\label{a1}
\end{eqnarray}
in the presence of a uniform external field $B$. We chose the next
nearest (nn) neighbour coupling $J_1$ to be antiferromagnetic ($J_1 >
0$) and use the notation $\alpha=J_2/J_1$. 
In the $ \alpha-\Delta $ - plane, we will primarily concentrate on the regime 
$\alpha \ge 0$, $-1 \le \Delta \le 1$.\\
The isotropic model with $\Delta=1$ and nnn
coupling $\alpha$ has been investigated by many authors
\cite{tonegawa1987,tonegawa1988,okamoto,farnell1994,gerhardt}. 
Most of these investigations
focussed on the transition \cite{haldane} from the 'spinfluid phase' $\alpha <
\alpha_c$ to the dimerphase $\alpha > \alpha_c$. The transition point
$\alpha_c=0.2411...$ has recently been determined with high precision
\cite{okamoto,eggert} by means of conformal field theory and 
renormalization group techniques. \\
The Hamiltonian with $\alpha=1/2$ , $\Delta=1$ has been studied first
by Majumdar and Gosh \cite{majumdar1969_1,majumdar1969_2}. 
They found that the 'dimer states'
\begin{eqnarray}
  |\psi> = \frac{1}{2^{N/4}} [1,2][3,4] ... [N-1,N] 
\label{a2} \\
  |\phi> = \frac{1}{2^{N/4}} [2,3][4,5] ... [N,1] 
\label{a3}
\end{eqnarray}
are eigenstates of $H(\alpha=1/2,\Delta=1,B=0)$.  Here
\begin{equation}
[i,i+1] = \frac{1}{\sqrt{2}} [\chi_+(i) \chi_-(i+1) - \chi_-(i) \chi_+(i+1)]
\label{a4}
\end{equation}
are nearest neighbour (nn) valence bond states with total spin zero, called
dimers. Van den Broek \cite{broek} proved that the dimer states are
indeed groundstates of the Hamiltonian at the 'Majumdar-Gosh' point
($\alpha=1/2, \Delta=1$).  
Affleck, Kennedy, Lieb and Tasaki \cite{affleck1987,affleck1988} 
were able to show that the
dimerstates are the only ground states and that there is a finite gap
to the first excited state. \\
Hamada, Kane, Nakagawa and Natsume \cite{hamada1988} discussed
uniformly distributed resonating valence bonds (UDRVB) in the
generalized railroad trestle model, 
which is equivalent to the isotropic linear Heisenberg chain with nn
and nnn interactions. They found that for negative $J_1$ and $J_2 = -
1/4 J_1$ the UDRVB is the ground state which is degenerate with the
fully magnetized state with total spin $S=S_z=N/2$. As we will show
later this phenomenon
also occurs for positive values of 
$J_1$ if the parameters $\alpha$ and $\Delta$ are properly
chosen in the Hamiltonian (\ref{a1}).  \\
Shastry and Sutherland discussed the frustrated model with
differen interaction strengths in $x,y$ and $z$ direction \cite{shastry}. 
The critical properties of the anisotropic model ($\Delta \ne 1$) in
the absence of an external field $B$ have been elaborated by
Nomura and Okamoto \cite{nomura}. They confirmed that this model and
the quantum sine-Gordon model belong to the same universality class. 
Tonegawa and Harada \cite{tonegawa1989} have studied Hamiltonian (\ref{a1}) with
ferromagnetic nn and antiferromagnetic nnn interactions for positive
$\Delta$. \\
The dimerstates (\ref{a2}), (\ref{a3})  are eigenstates 
of the anisotropic model along
the whole line $\alpha=\frac{1}{2}$ , $-\infty < \Delta < \infty$ as
will be shown explicitly in section 2. \\
However, the eigenvalues 
\begin{equation}
E_D(\alpha=\frac{1}{2},\Delta) = -N (\frac{1}{4}+\frac{\Delta}{8})
\label{a5}
\end{equation}
are groundstate energies only for $\Delta > -\frac{1}{2}$. For $\Delta
< -\frac{1}{2}$ the groundstate $|F\pm>$ with energy
\begin{equation}
E_F(\alpha=\frac{1}{2}, \Delta) = -\frac{3}{8} \Delta N
\label{a6}
\end{equation}
is found in the ferromagnetic  sector where all spins are
down or up.
This is a first hint, that the  model (\ref{a1}) is particularly
suited to study the transition from antiferromagnetism to
ferromagnetism. \\

The outline of the paper is as follows: In section 2 we report on the
quantum numbers and the finite size effects of the groundstates as
they depend on $\alpha, \Delta$ and the magnetization
$M=S_z/N$. Section 3 is devoted to an analysis of the magnetic
properties of the model (\ref{a1}). Three phases can be found in the
$\alpha-\Delta$-plane: the ferromagnetic, the antiferromagnetic and the metamagnetic
phase. \\
The Hamiltonians with $\alpha=0$, $\Delta=-1$ and $\alpha=0.5$,
$\Delta=-0.5$ are special in the sense that the groundstate is highly
degenerate --- namely with respect to $S_z=0, \pm 1, \pm2, ... ,\pm
N/2$. This feature is discussed in section 4.

\section{Quantum numbers and finite size effects in the groundstate}
Let us start with the groundstate properties of the Hamiltonian
(\ref{a1}) in the strip
\begin{equation}
0 \le \alpha \le \frac{1}{2} .
\label{b1}
\end{equation}
In the absence  of a magnetic field, the groundstate is found in the
sector with total spin $S_z=0$  and momentum $p_0$  ($p_0=0, N=2n , n$
even, $p_0=\pi, N=2n, n$ odd). We obtain the ground state energies
$E(S_z,\alpha,\Delta,N)$ on
finite systems up to $N=30$ through a direct Lanczos diagonalization,
making use of the translational invariance of Hamiltonian (\ref{a1}). 
This reduces the dimension of the Hilbert space approximately by a
factor of $N$. We choose a set of base states as proposed in
Ref. \onlinecite{takahashi} by Takahashi. This choice results 
in a real Hamiltonian matrix even for momenta $p \ne 0,\pi$. 
Hence we obtain the proper groundstate of the model even if its
momentum is not $p=0$ or $p=\pi$. \\
Along the line $\alpha=\frac{1}{2}, \Delta > -\frac{1}{2}$ the
groundstate is twofold 
degenerate; the two states are just the dimer states (\ref{a2}),
(\ref{a3}). 
The proof of this statement follows the arguments of van den Broek
\cite{broek}. The Hamiltonian
\begin{equation}
H(\alpha=\frac{1}{2},\Delta,B=0) = \sum_i H(i,i+1,i+2,\Delta)
\label{b2}
\end{equation}
is expressed in terms of three spin Hamiltonians:
\begin{eqnarray}
 H(i,i+1,i+2,\Delta) &=& \frac{1}{4} 
(S^+_{i}S^-_{i+1}+S^-_{i}S^+_{i+1}+S^+_{i+1}S^-_{i+2}+S^-_{i+1}S^+_{i+2}
 + S^+_{i}S^-_{i+2}+S^-_{i}S^+_{i+2}) 
\nonumber  \\
  & & + \frac{\Delta}{2} (S^z_{i}S^z_{i+1} + S^z_{i+1}S^z_{i+2} + S^z_{i}S^z_{i+2})
\label{b3}
\end{eqnarray}
The dimer states (\ref{a2}), (\ref{a3}) 
turn out to be eigenstates of $H(i,i+1,i+2,\Delta)$
with eigenvalue $\epsilon_0(\Delta)=-(\frac{1}{4}+\frac{\Delta}{8})$. One can easily
prove that this is the lowest eigenvalue of the three spin Hamiltonian
for $\Delta > -\frac{1}{2}$. However, this is not the case for $\Delta <
-\frac{1}{2}$, where the lowest eigenvalue of the three spin Hamiltonian is given by
$\epsilon_1(\Delta)=-\frac{3\Delta}{8}$. The corresponding eigenstate
has all three spins pointing in the same direction (cf. (\ref{a6})).

Let us next turn to the finite size effects of the groundstate
energy. Within the strip $0 \le \alpha \le \frac{1}{2}, \Delta > -\frac{1}{2}$, they
turn out to be monotonically increasing with $N$. Finite size effects
vanish on the 'dimerline' $\alpha=\frac{1}{2}, \Delta \ge -\frac{1}{2}$, where the
groundstate is completely dimerized and degenerate. Right to the
dimerline, i.e. for 
$\alpha > \frac{1}{2}, \Delta \ge -\frac{1}{2}$, the groundstate
properties  -- with respect to its momentum quantum numbers -- change
and the monotonic behaviour of the finite size
effects is lost. 
\\
In the presence of a uniform magnetic field $B$ with
magnetization $M(B)=S_z/N$ the ground state of the isotropic model
$(\Delta=1)$ is found in the sector with
total spin $S_z=S$. A rule for the momenta $p_s$ of these states can be deduced
from Marshall's sign rule 
\cite{marshall}:
\begin{eqnarray}
p_s = 0 \quad \mbox{ for } \quad  2S + N = 4n  \nonumber \\ 
p_s = \pi \quad \mbox{ for } \quad 2S + N = 4n + 2  .
\label{b4}
\end{eqnarray}
This rule has been proven \cite{tonegawa1987,gerhardt} 
to be correct in the unfrustrated case,
however it turned out to be valid in a larger M-dependent domain in
the $\alpha-\Delta-$ plane. E.g. in the isotropic case 
($\Delta=1$) we found \cite{gerhardt}
that the momentum rule (\ref{b4}) is satisfied for $\alpha <
\alpha_0(M)$ i.e. below some curve $\alpha_0(M)$, which starts at the
Majumdar Gosh point 
\begin{equation}
\alpha_0(M=0,\Delta=1)=\frac{1}{2}
\label{b5}
\end{equation} 
and ends at
\begin{equation}
\alpha_0(M=\frac{1}{2},\Delta=1)=\frac{1}{4}.
\label{b6} 
\end{equation}
The groundstate is degenerate along the curve
$\alpha_0(M,\Delta)$. The two states differ in their momenta; the first one
follows (\ref{b4}). A rule for the momentum of the second
state has not yet been found. \\
In Fig. 1 we have plotted numerical results of the curves
$\alpha_0(M,\Delta)$ for various values of the anisotropy parameter
$\Delta=1.0, 0.4, 0.1, -0.2$. The data for $\alpha_0(M,\Delta)$ mark
those points in the $\alpha-M$-plane, where the groundstates with
energy $\epsilon(M,\alpha_0(M,\Delta),\Delta,N)$, $N=12,...,18$ are
twofold degenerate. Finite size effects of $\alpha_0(M,\Delta)$ are
visible at $M=1/4$ and $M=1/6, 1/3$ where systems of size $N=12,16$
and $N=12,18$, respectively, are realized. In spite of the finite size
effects we think that the finite system results shown in Fig. 1
reproduce the qualitative features of the curves $\alpha_0(M,\Delta)$
in the thermodynamical limit:
\begin{itemize}
\item all curves start and end at the points (\ref{b5}) and
(\ref{b6}). 
\item $\alpha_0(M,\Delta=1)$ has a pronounced maximum  around
$M=0.2$ with rather large finite size effects. For decreasing values
of $\Delta$ the height of the maximum is reduced  and its position is
shifted to smaller values of $M$.
\end{itemize}
Beyond the curve $\alpha_0(M,\Delta)$  --- i.e. for
$\alpha>\alpha_0(M,\Delta)$ --- the groundstate momenta  deviate from
the rule (\ref{b4}) and we therefore expect a change in the
groundstate properties.

\section{The phase diagram in the presence of a magnetic field}

In this section we will present numerical results for the groundstate
energy per site
$
\epsilon(M,\alpha,\Delta,N) = \frac{1}{N}  E(S_z,\alpha, \Delta, N) .
$
We are in particular interested in the changes of the M-dependence of
these energies
with $\alpha$ and $\Delta$ since they indicate a change in the
groundstate ordering. The following situations have been found:
\begin{itemize}
\item Ferromagnetic phase: \\
Here the free energy 
\begin{equation}
f(M,\alpha,\Delta) = \epsilon(M,\alpha,\Delta) - B M 
\label{c1}
\end{equation}
is minimized
by the states $|F\pm>$, where all spins are up ($+$) or down ($-$),
respectively. It turns out that the boundary of the ferromagnetic
phase $\Delta<\Delta_f(\alpha)$ is characterized by the degeneracy 
\begin{equation}
\epsilon(M=0,\alpha,\Delta_f(\alpha),N) = 
\epsilon(M=\frac{1}{2},\alpha,\Delta_f(\alpha),N) 
\label{c2}
\end{equation}
of the lowest energy eigenvalues in the sectors with $S_z=0$ and
$S_z=N/2$. 
\item Antiferromagnetic phase: \\
The minimum of the free energy is found for $0 \le M \le \frac{1}{2} $
at
\begin{eqnarray}
\frac{d \epsilon}{d M} - B = 0 
\label{c3} \\
\frac{d^2 \epsilon}{d M^2} > 0 
\label{c4}
\end{eqnarray}
This means that $\epsilon(M,\alpha,\Delta,N)$ is monotonically
increasing and convex for $0 \le M \le \frac{1}{2}$. 
The saturating field
\begin{equation}
\frac{d \epsilon}{d M} |_{M=\frac{1}{2}} =
B(M=\frac{1}{2},\alpha,\Delta) ,
\label{c5}
\end{equation}
which is needed to align all spins in the system, can be computed from
the one magnon states:
\begin{equation}
|p,S_z=\frac{N}{2}-1> = \frac{1}{\sqrt{N}} \sum_x e^{ipx} |x>  ,
\label{c6}
\end{equation}
where $|x>$ denotes the state with one spin down at site $x$
and all other spins up. The energy of this state is
\begin{equation}
 E(S_z=\frac{N}{2}-1,\alpha,\Delta,N) = \cos p + \alpha \cos 2p +
\Delta(1-\alpha)(\frac{N}{4}-1)
\label{c7}
\end{equation}
and the groundstate energy is found by minimization with respect to $p$. 
For $0 \le \alpha \le 1/4$ the minimum is found at $p=\pi$ and the
saturating field is 
\begin{equation}
B(M=\frac{1}{2},\alpha,\Delta)=\Delta(1+\alpha) + (1-\alpha)  .
\label{c8}
\end{equation}
For $\frac{1}{4} < \alpha \le \frac{1}{2}$, the minimum energy
(\ref{c7}) is found for 
\begin{equation}
\cos p = -\frac{1}{4\alpha}
\label{c9}
\end{equation}
which yields for the saturating field 
\begin{equation}
B(M=\frac{1}{2},\alpha,\Delta) = \Delta (1+\alpha) + \alpha + \frac{1}{8
\alpha}  .
\label{c10}
\end{equation}
The boundary $\Delta_a(\alpha)$ of the antiferromagnetic phase 
$\Delta > \Delta_a(\alpha)$ is characterized by the condition 
\begin{equation}
\frac{d^2 \epsilon}{d M^2} (M,\alpha,\Delta_a(\alpha),N) |_{M=\frac{1}{2}}
= 0
\label{c11}
\end{equation}
i.e. the convexity condition is lost for $\Delta < \Delta_a(\alpha)$.
\item Metamagnetic phase: \\
Between the ferromagnetic and the antiferromagnetic phase
\begin{equation}
\Delta_f(\alpha) < \Delta < \Delta_a(\alpha) 
\label{c12}
\end{equation}
we find a metamagnetic phase, which  is characterized be a zero in the second
derivative:
\begin{eqnarray}
\frac{d^2 \epsilon}{d M^2} > 0, \hspace{1cm}  0 < M < M_c(\alpha,\Delta) 
\label{c13} \\
\frac{d^2 \epsilon}{d M^2} (M,\alpha,\Delta,N)|_{M=M_c} =  0
\label{c14}
\end{eqnarray}
The minimum of the free energy is found for 
\begin{equation}
\frac{d \epsilon}{d M} = B \quad \mbox{for} \quad 0<M<M_c(\alpha,\Delta)
\label{c15}
\end{equation}
and at 
\begin{equation}
M=\frac{1}{2} \quad \mbox{for} \quad B > B_c=\frac{d \epsilon}{d M} |_{M=M_c}.
\label{c16}
\end{equation} 
Therefore, in this metamagnetic phase 
we have a discontinuity at $B_c(\alpha,\Delta)$ where the
magnetization curve jumps from $M=M_c(\alpha,\Delta)$ to $M=\frac{1}{2}$.
$B_c(\alpha,\Delta)$ decreases, 
if one crosses the metamagnetic phase coming
from the antiferromagnetic phase and moving towards the ferromagnetic
phase. An example will be given below. \\
For small magnetic fields $0<B<B_c$ the system looks
antiferromagnetic, for $B>B_c$ ferromagnetic. 
\end{itemize}
For the determination of the phase boundaries $\Delta_f(\alpha)$ and
$\Delta_a(\alpha)$ of the 
ferromagnetic and antiferromagnetic phase, we have first computed the
lowest energy  densities $\epsilon(M,\alpha,\Delta,N)$  
as they depend on the magnetization $M$ and the
parameters $\alpha$ and $\Delta$. As an example we show in Fig. 2a) the
evolution of the $M$-dependence for $\Delta=-0.6, \alpha=0.0, ... ,
0.40$ on a system with $N=16$ sites. One clearly observes the three
phases discussed above. For small values of $\alpha$, $\epsilon(M,\alpha,\Delta)$ is
monotonically increasing and convex. Here, we are in the
antiferromagnetic phase. At $\alpha=0.15  $ the second derivative
(\ref{c14}) vanishes
first at $M_c=\frac{1}{2}$. 
The metamagnetic phase extends from $\alpha=0.15$ to $\alpha = 0.305$, 
where the degeneracy (\ref{c2}) shows up. The behaviour of the
corresponding magnetization curves can be seen in Fig. 2b). We
obtain these magnetization curves by applying the method of Bonner
and Fisher \cite{bonner} to our finite system results. For 
$\alpha > 0.305$ we then enter the ferromagnetic phase. The resulting phase
diagram in the $\alpha-\Delta-$ plane is shown in Fig. 3 . The
numerical evaluation of (\ref{c2}) on finite systems does not show a
significant finite size dependence. In other words, the determination
of the phase boundary $\Delta_f(\alpha)$ is well under control. \\
The determination of the second phase boundary $\Delta_a(\alpha)$ from 
(\ref{c11}) turned out to be much more difficult. We 
numerically calculated $\epsilon(M=1/2-2/N,\alpha,\Delta,N)$ --
i.e. the lowest eigenvalue in the sector with two spins flipped --  on
rather large systems with  
$N=20,30,40,50$  and looked for a zero in the second derivative:
\begin{equation}
 \epsilon(M=\frac{1}{2}-\frac{2}{N},\alpha,\Delta,N) +
\epsilon(M=\frac{1}{2},\alpha,\Delta,N) - 
2\epsilon(M=\frac{1}{2}-\frac{1}{N},\alpha,\Delta,N) = 0
\label{c17}
\end{equation}
The resulting $\Delta_a(\alpha)$ suffers under finite size effects
particularly in the vicinity of the point $\alpha=0.5$,
$\Delta=-0.5$. The curve $\Delta_a(\alpha)$ plotted in the phase
diagram (Fig. 3) represents the result of (\ref{c17}) for the largest
system size $N=50$.

The points $\alpha=0, \Delta=-1$ and $\alpha=\frac{1}{2}, \Delta=-\frac{1}{2}$ are
special in the sense, that the boundaries $\Delta_a(\alpha),
\Delta_f(\alpha)$ for the 
antiferromagnetic and ferromagnetic phase meet. Therefore, we have no
metamagnetic phase between the ferro- and antiferromagnetic phase at these
points. This can also be clearly seen in Figs. 4a and 4b, where we
have plotted the $M$ and $\Delta$ dependence of
$\epsilon(M,\alpha=0,\Delta)$  and $\epsilon(M,\alpha=1/2,\Delta)$,
respectively.  These energies turn out to be convex for $\Delta > -1$,
$(\alpha=0)$ and $\Delta > -\frac{1}{2}$, $(\alpha=0.5)$ and concave
for $\Delta < -1$, 
$(\alpha=0)$, $\Delta  < -\frac{1}{2}, \quad (\alpha=0.5)$, respectively. At $\Delta=-1,
\alpha=0$ and $\Delta=-\frac{1}{2}, \alpha=\frac{1}{2}$, the groundstate energies
$\epsilon(M,\alpha,\Delta,N)$ are all degenerate with respect to
$M$. This feature will be 
investigated in the next section.

\section{Groundstate degeneracy at the points where the phase
boundaries meet}
According to (\ref{c2})  
the phase boundary $\Delta_f(\alpha)$ of the ferromagnetic phase is
defined by the degeneracy of two eigenstates with total spin $S_z=0$
and $S_z=\frac{N}{2}$ , respectively.
At the points $\Delta=-1, \alpha=0$ and $\Delta=-\frac{1}{2}, \alpha=\frac{1}{2}$
where $\Delta_f(\alpha)=\Delta_a(\alpha)$, a much larger degeneracy
of the groundstate with 
respect to all values of $S_z=0, ...  ,\frac{N}{2}$ occurs. We are
going to show now, that the n-magnonstates $S_+(p)^n |F->$ --- obtained
by n-fold application of the rising operator
\begin{equation}
S_+(p) = \sum_l e^{ipl} S^+_l
\label{d1}
\end{equation}
on the ferromagnetic state $|F->$ --- are eigenstates of the Hamiltonian
\begin{equation}
H(\alpha,\Delta,B) S_+(p)^n |F-> = E_F S_+(p)^n |F->
\label{d2}
\end{equation}
for
\begin{eqnarray}
\mbox{a)} \quad & & \alpha=0, \quad \Delta=-1, \quad p=\pi 
\label{d3} \\
\mbox{b)} \quad & & -\infty < \alpha < \infty, \quad \Delta=-\frac{1}{2}, \quad
p=\frac{2\pi}{3}, \quad p=\frac{4\pi}{3} .
\label{d4}
\end{eqnarray}
Here
\begin{equation}
E_F = \frac{N}{4} (\Delta+\Delta \cdot \alpha)
\label{d5}
\end{equation}
is the energy of the ferromagnetic state. 
For the proof of (\ref{d2}) we start from
the commutation relations:
\begin{equation}
 [H(\alpha,\Delta,B),S_+(p)] = \sum_{j=1}^2 
(a_j \sum_l S^+_l S^z_{l+j} e^{ipl} + b_j \sum_l  S^z_{l} S^+_{l+j} e^{ipl}) 
\label{d6}
\end{equation}
where 
\begin{eqnarray}
a_1 = -e^{ip} + \Delta,  \quad   b_1 = -1+\Delta e^{ip} 
\nonumber \\
a_2 = \alpha (-e^{2ip} + \Delta) \quad  b_2 = \alpha(-1+\Delta e^{2ip})
\label{d7}
\end{eqnarray}
and 
\begin{equation}
[[H(\alpha,\Delta),S_+(p)], S_+(p)] = \sum_{j=1}^2 c_j \sum_l
S^+_l S^+_{l+j} e^{2ipl}
\label{d8}
\end{equation}
where 
\begin{eqnarray}
c_1= 2 e^{ip} (\Delta - \cos p)   \nonumber \\   
c_2= 2 \alpha e^{2ip}  (\Delta - \cos 2p)       
\label{d9}
\end{eqnarray}
All further commutators with $S_+(p)$ vanish identically.
Application of (\ref{d6}) and (\ref{d8}) onto the ferromagnetic state $|F->$ yields:
\begin{equation}
[H(\alpha,\Delta),S_+(p)] |F-> = -2(\Delta - \cos p +
\alpha(\Delta - \cos 2p)) |p>
\label{d10}
\end{equation}
where $|p>$ is the one-magnon state (\ref{c6}). 
Similarly one finds 
\begin{eqnarray}
[[H(\alpha,\Delta),S_+(p)], S_+(p)] |F-> & = & 
2 e^{ip} (\Delta - \cos p) |2p,1>  \nonumber \\
& & + 2 e^{2ip} \alpha (\Delta - \cos 2p) |2p,2>
\label{d11}
\end{eqnarray}
where 
\begin{equation}
|2p,j> = \sum_l e^{2ipl} |l,l+j>
\label{d12}
\end{equation}
are two magnon states with two spins up at sites $l$ and $l+j$. Note,
that the right hand sides of (\ref{d10}) and (\ref{d11}) 
vanish for the two cases listed in (\ref{d3})
and (\ref{d4}) , respectively. The first point (\ref{d3}) marks the transition from
antiferromagnetism to ferromagnetism in the nearest neighbour $XXZ$
model. Indeed, the rising operator $S_+(\pi)$ commutes with the
Hamiltonian $H(\alpha=0,\Delta=-1)$. \\
Along the line (\ref{d4}), the endpoint of the dimerline ($\alpha=0.5,
\Delta=-0.5$)
is of special interest. Here the eigenvalues (\ref{a5}) of
the dimerstates (\ref{a2}), (\ref{a3}) are degenerate with the energy
(\ref{d5}) of the ferromagnetic states. This also implies that the
n-magnonstates (\ref{d2}) are groundstates for $\alpha=\frac{1}{2}$,
which explains the degeneracy found in Fig. 4 a for
$\alpha=-0.5$. \\
In Fig. 5 we have plotted the groundstate energies
$\epsilon(M,\alpha,\Delta=-0.5)$ on a ring with 18 sites along the line
(\ref{d4}), where the degeneracy of the n-magnon states (\ref{d2}) has
been proven. For $\alpha < 0.5$ the groundstate energies
$\epsilon(M,\alpha,\Delta=-0.5)$ are monotonically increasing with $M$;
i.e. the corresponding groundstates cannot be identified with the
degenerate n-magnonstates (\ref{d2}). The groundstate energies
$\epsilon(M,\alpha,\Delta=-0.5)$ meet each other for all $M$ at
$\alpha=0.5$ and stay very close together  in the interval $0.5 < \alpha <
0.6$. This leads to the narrow width of the metamagnetic phase in
Fig. 3 for $0.5 < \alpha < 0.6$.
For $\alpha>0.6$ the quasi-degeneracy with respect to M is lifted again. 

\section{Discussion and conclusion}
Metamagnetism denotes a mixed phase between ferromagnetism and
antiferromagnetism, which has been observed in various substances like
e.g. $Fe_xMn_{1-x}TiO_3$. The characteristic signal is a rapid
increase (or discontinuity) in the magnetization curve, if the
external field exceeds a critical value $B_c$. For $B>B_c$ the
substance is almost fully magnetized. In this paper we have shown that
the phenomenon of metamagnetism  can be observed in the one
dimensional spin-$1/2$ XXZ-model with next to nearest neighbour
coupling $\alpha$ and anisotropy $\Delta$. 
The phase diagram in the $\alpha-\Delta$-plane (Fig. 3) contains three
regimes: the antiferromagnetic one with $\Delta>\Delta_a(\alpha)$, the
ferromagnetic one with $\Delta<\Delta_f(\alpha)$ and the metamagnetic
one in between $\Delta_a(\alpha) \ge \Delta \ge \Delta_f(\alpha)$. The
metamagnetic phase shrinks to zero at $\alpha=0$ and $\alpha=0.5$,
where $\Delta_a(\alpha)=\Delta_f(\alpha)$. At these points there is a
direct transition from antiferromagnetism to ferromagnetism and the
groundstate turns out to be highly degenerate --- namely with respect
to $S_z=0,1,2,...,N/2$. These states can be identified with
n-magnon states.

\newpage
\centerline{\bf Figure Captions}
\vspace*{2cm}
\noindent
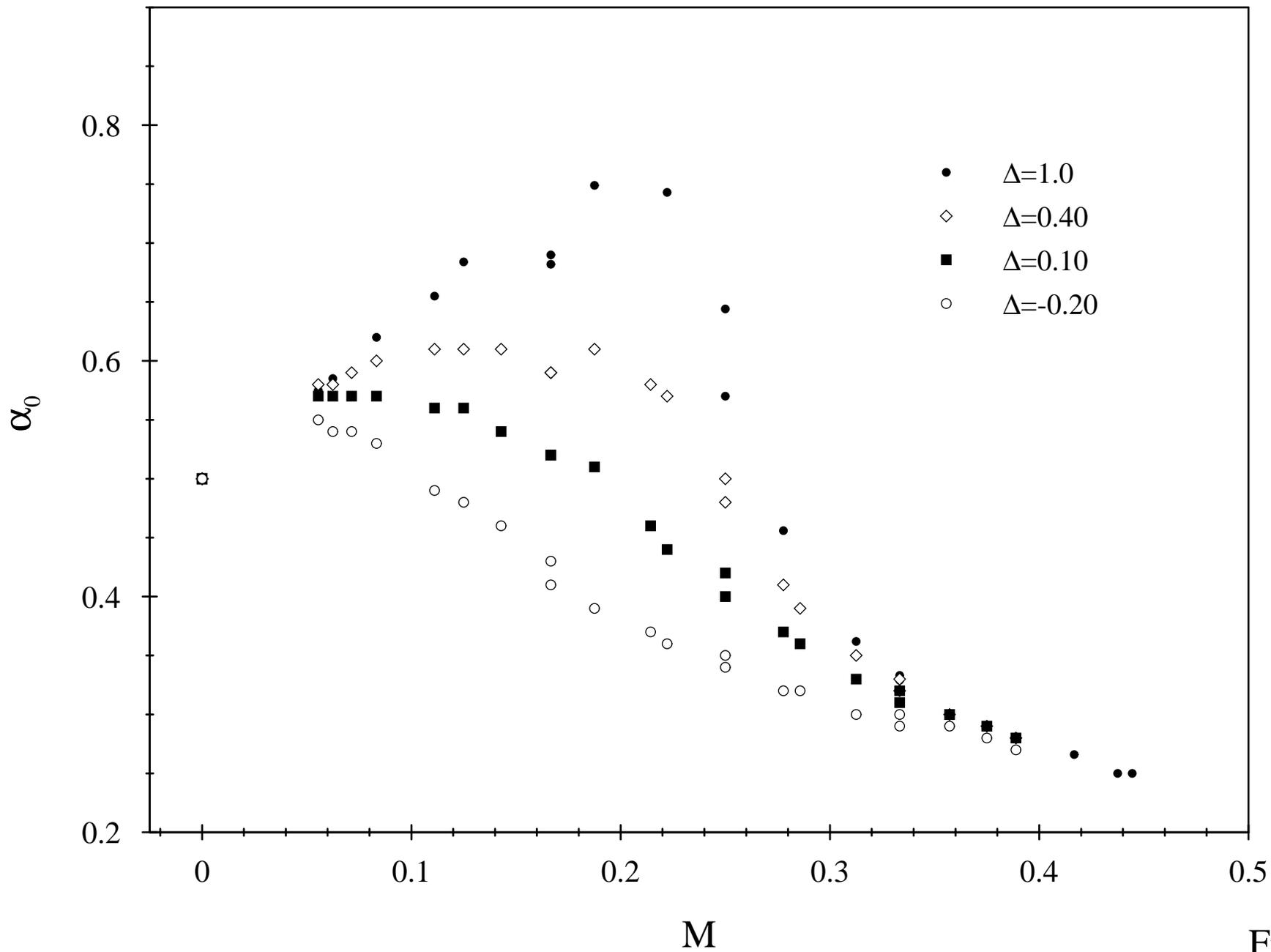
FIG. 1. The groundstates of the Hamiltonian (\ref{a1}) are degenerate along
the curves $\alpha_0(M,\Delta)$. The numerical data points were
obtained on finite system calculations with $N=12,14,16,18$ and are
shown for $\Delta=1.0, 0.4, 0.1, -0.2$. \\[10mm]
FIG. 2. (a) $M$-dependence of the groundstate energy per site
$\epsilon(M,\alpha,\Delta,N)$ with $N=16$, $\Delta=-0.6$, and
$\alpha=0.0, 0.1, ..., 0.4$. 

(b) Magnetization curves $M(B,\alpha,\Delta)$ for $\Delta=-0.6$ as they
follow from the groundstate energy per site
$\epsilon(M,\alpha,\Delta,N)$ shown in Fig. 2a . In the metamagnetic
phase (e.g. at $\alpha=0.2, 0.25$, $\Delta=-0.6$) there is a critical
field $B_c(\alpha,\Delta)$, where the system jumps into the fully
magnetic state with $M=1/2$. \\[10mm]
FIG. 3. The phase diagramm in the $\alpha-\Delta$-plane, as it follows from
the numerical evaluation of (\ref{c2}) ($N=18$) and (\ref{c10})
($N=50$). The metamagnetic domain $\Delta_f(\alpha) < \Delta <
\Delta_a(\alpha)$ is hatched. \\[10mm]
FIG. 4. Groundstate energies $\epsilon(M,\alpha,\Delta,N=20,24)$ along the
lines \\
(a) $\alpha=0.5, \quad \Delta=1.0, 0.6, ..., -1.0 $ ,
(b) $\alpha=0.0, \quad \Delta=1.0,  ..., -1.8 $ .\\
The groundstate degeneracy is clearly visible for  $\alpha=0.5, \quad
\Delta=-0.5$ and $\alpha=0, \quad \Delta=-1$, where the direct
transition from antiferromagnetism to ferromagnetism  occurs. \\[10mm]
FIG. 5. $\alpha$ and $M$ dependence of the groundstate energies
$\epsilon(M,\Delta,N=18)$ along the line $\Delta=-0.5$.

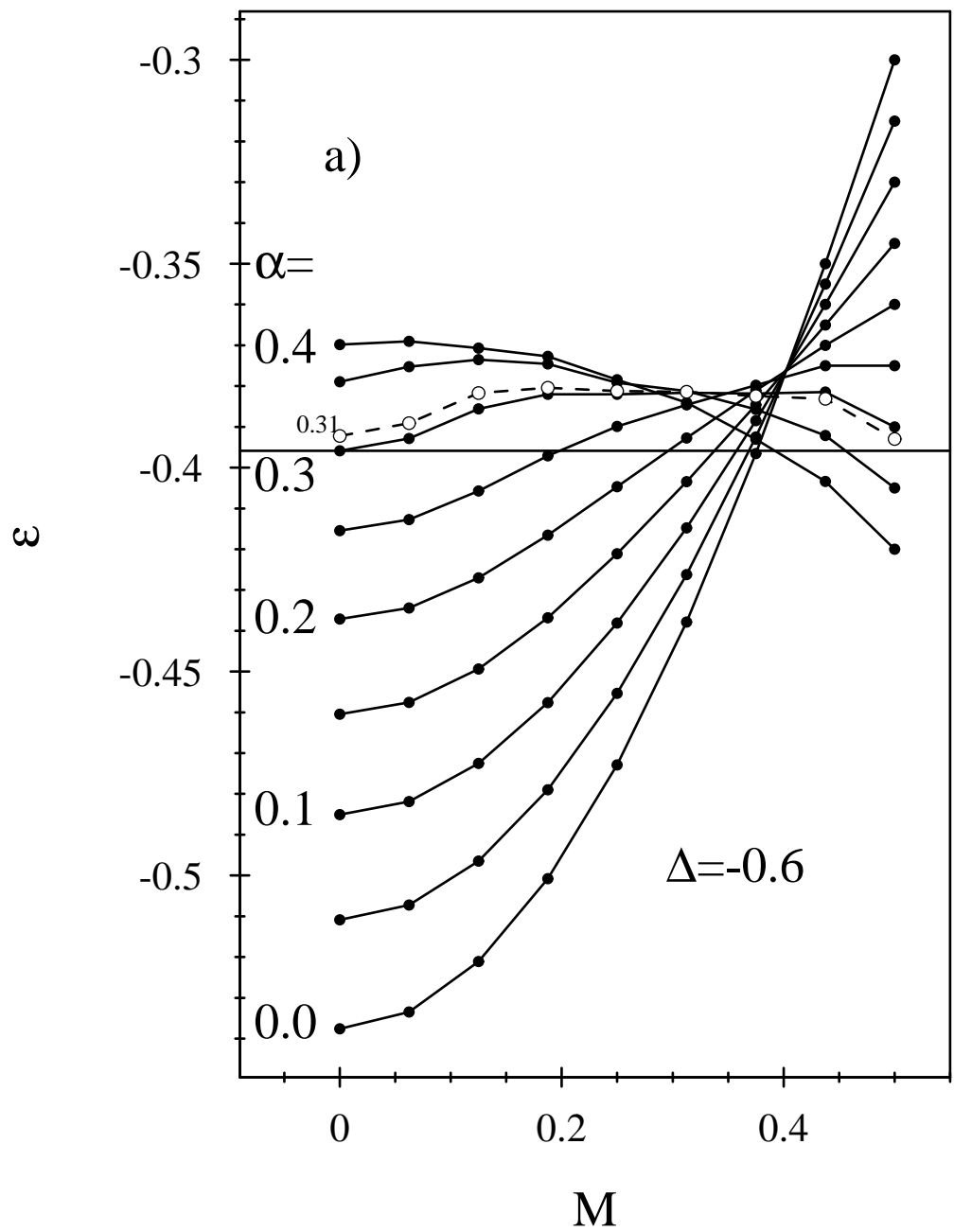
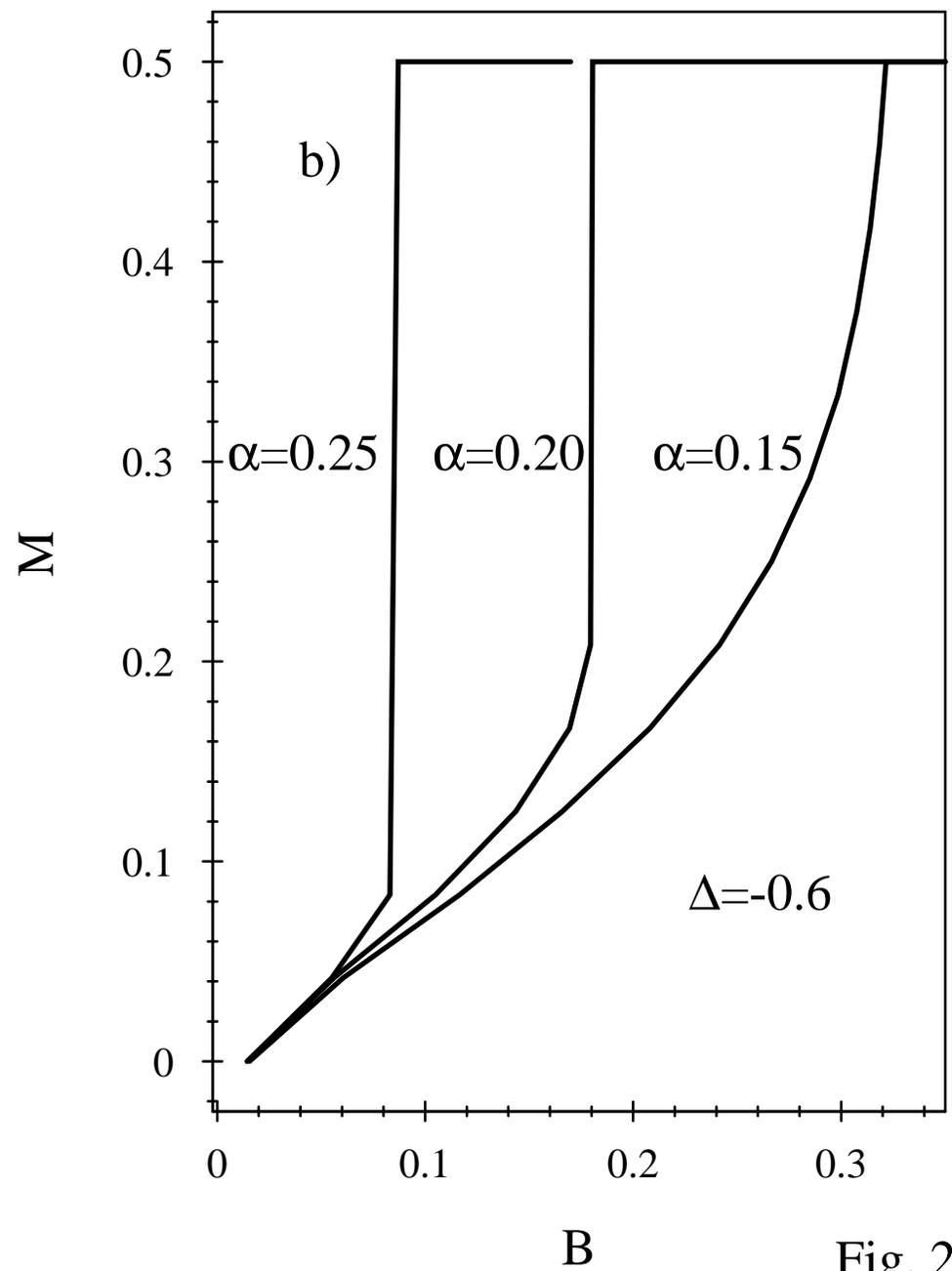
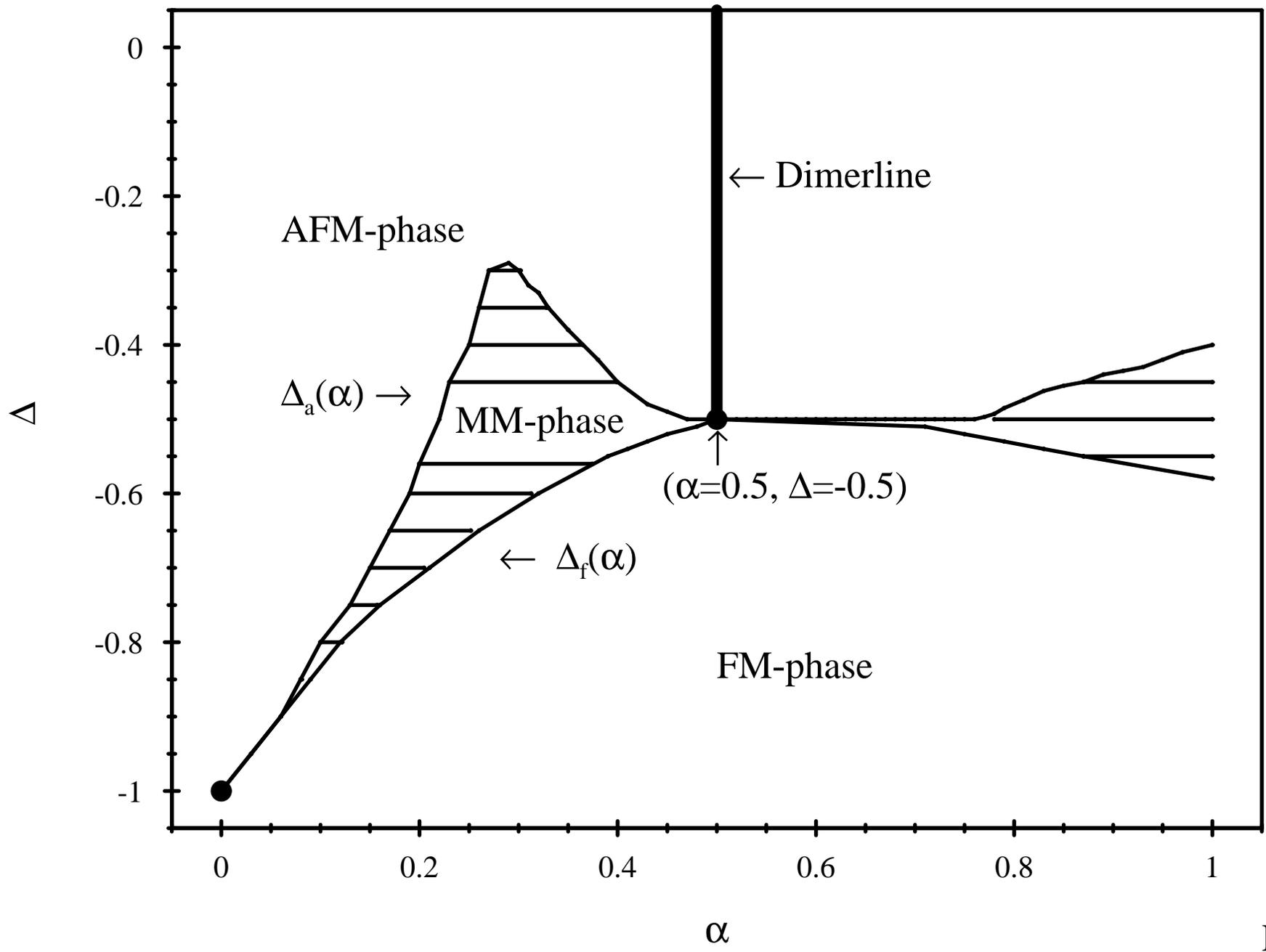
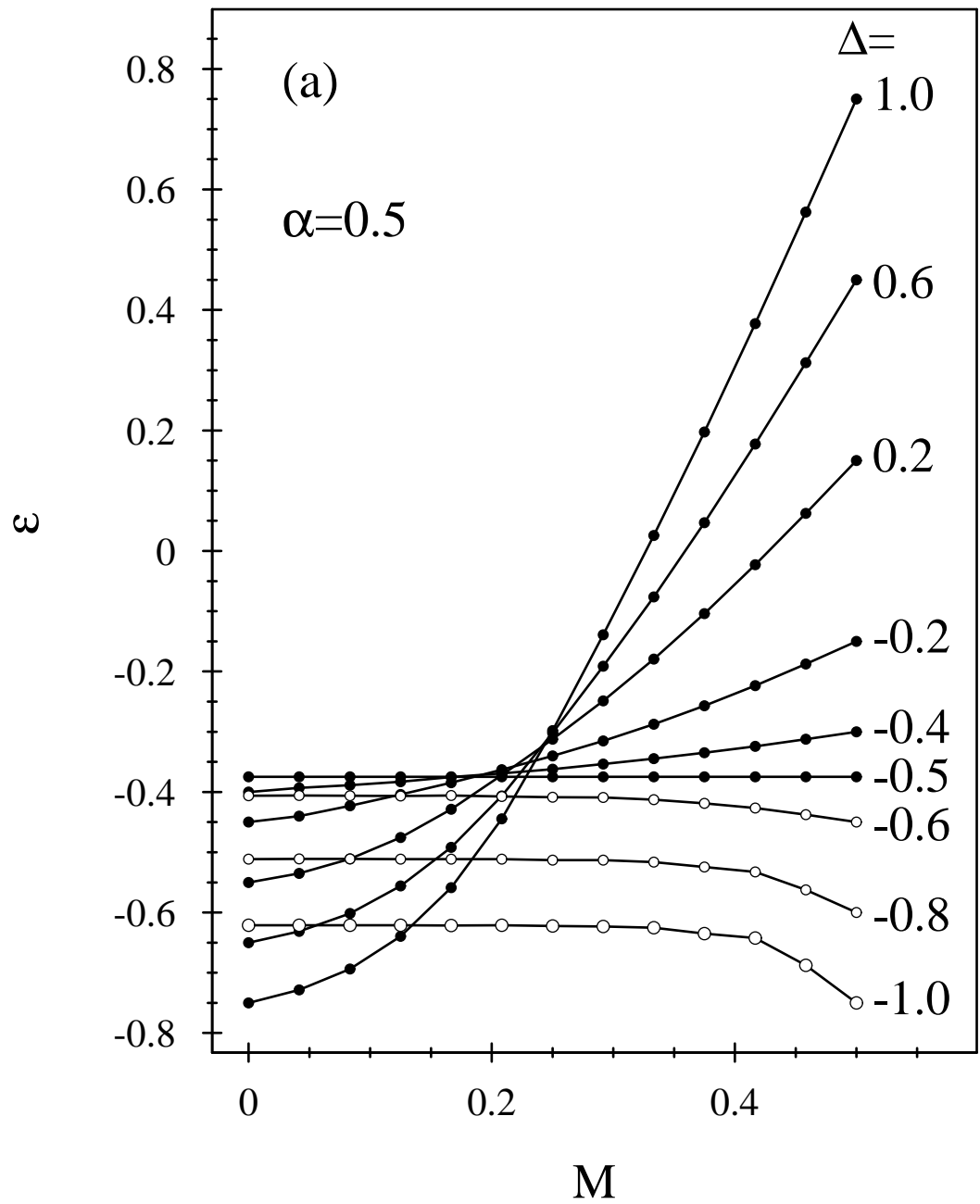
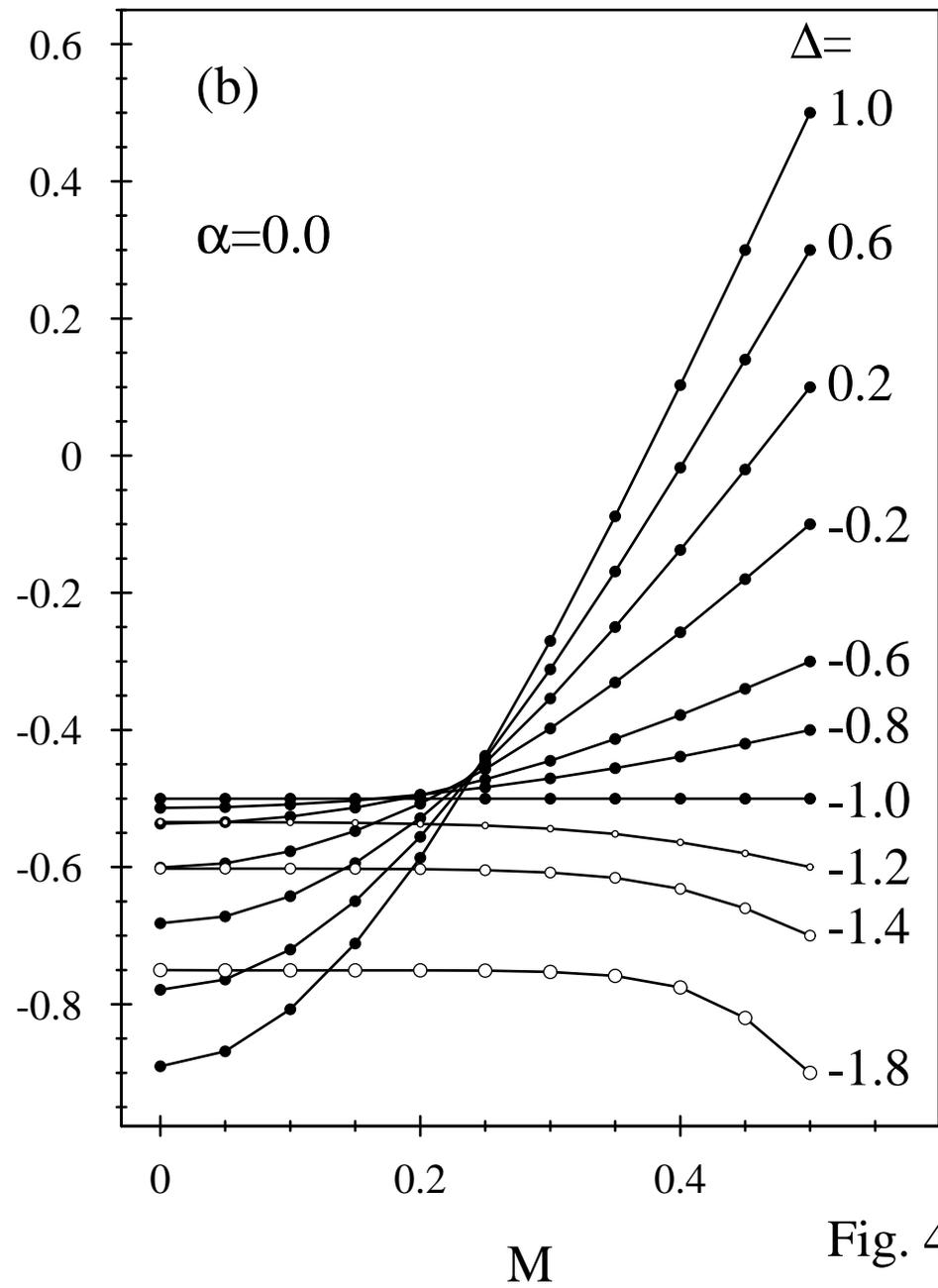
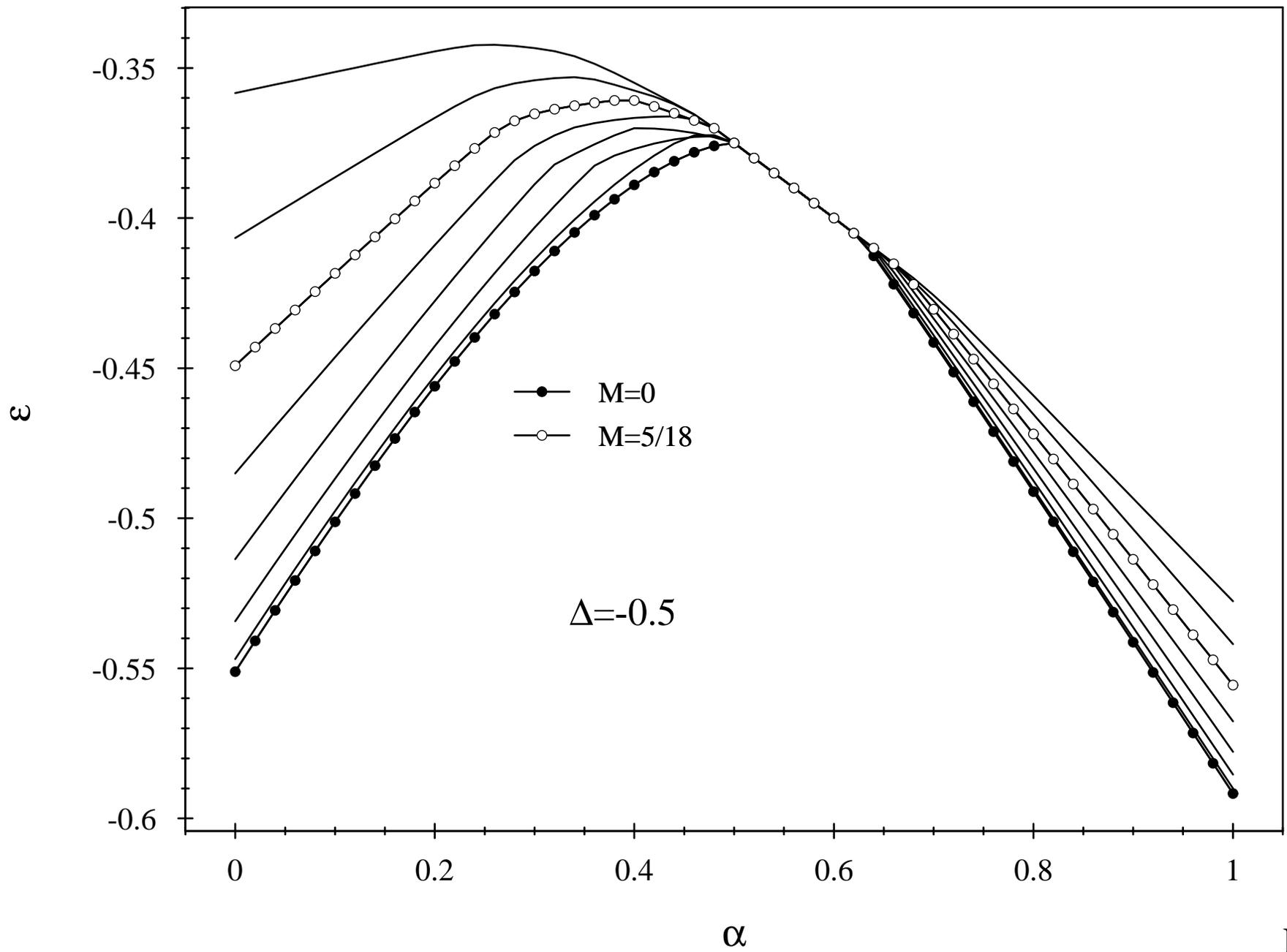
\end{document}